\documentstyle[a4wide,epsf,11pt,titlepage]{article}

\newcounter{nref}
\newcommand{\bbib}{%
  \renewcommand{\refname}{\large\bf References}%
  \begin{thebibliography}{99}%
  \setcounter{enumiv}{\thenref}}
\newcommand{\ebib}{%
  \end{thebibliography}%
  \setcounter{nref}{\arabic{enumiv}}}
\newcommand{\head}[3]{%
  \setcounter{nref}{0}%
  \thispagestyle{empty}%
  \section*{\LARGE\bf #1}%
\setcounter{section}{-1}
  \stepcounter{section}%
  \addcontentsline{toc}{section}{#1}%
  \large\itshape%
  #2\\\vspace{0.1pt}\\%
  #3%
  \normalsize\upshape%
  \bigskip}

\def\deg{{$^{\circ}$}}
\def\bd17{\mbox{BD +17\deg 3248}}
\def\third{{3$^{\rm rd}$}}
\begin{document}


\head{The r-Process and Chronometers}
     {John J. Cowan\ $^1$,  Christopher Sneden\ $^2$ and James W. Truran\ $^3$} 
     {$^1$ Department of Physics and Astronomy, University of Oklahoma,
Norman, OK 73019\\
     $^2$ Department of Astronomy and McDonald Observatory,
University of Texas, Austin, TX 78712\\
      $^3$ Department of Astronomy \& Astrophysics, University of Chicago,
Chicago, IL 60637
}

\section{Introduction}

We have been examining  the abundance distributions in metal-poor Galactic halo
stars trying to identify, and understand,
 the signatures of the slow- and rapid-neutron 
capture processes ({\it i.e.}, the $s$- and the $r$-process).
Detailed studies have been made of the stars CS~22892--052 \cite{sne96,sne00} 
and HD~115444 \cite{wes00}.
In our latest work we have added to this list the 
metal-poor ([Fe/H] = --2.0) halo star
\bd17 \cite{cow02}.

\section{Observations  and Abundance Trends in \bd17}

Using both ground-based (Keck and McDonald) and space-based 
(Hubble Space Telescope,
HST) observations we have obtained high resolution, high signal-to-noise
spectra of \bd17. The element gold has been  detected in \bd17 
as shown in Figure 1.
This HST STIS observation shows a comparison between \bd17 and HD 122563.
While the spectra appear somewhat similar in this region, 
this is due to the
complex set of absorption lines that contribute to the total feature
near this wavelength.
Specifically, there is an OH line near  the  2675.94~\AA\ gold line.
This results in some blending of the gold feature for \bd17. 
The OH absorption, however, is much stronger throughout the UV spectrum of 
HD~122563 and is solely repsonsible for the strong line at 2676.0~\AA\ 
in this star. We note that the this gold detection in \bd17 is the first 
in any metal-poor halo stars.

We show in Figure 2 the entire abundance distribution of \bd17. 
This includes the element uranium. 
We have detected  a weak line at 
3859.60~\AA\ in the spectrum of \bd17\ 
that we tentatively identify with this element. 
This would be  only the second such detection, after Cayrel 
{\it et al.}\cite{cay01}, 
in any metal-poor halo stars.

We note in Figure 2 that the heavier neutron-capture elements, Z$\ge$ 56,
including now the \third\ $r$-process peak elements, all fall on the scaled
solar system $r$-process distribution. This is the same pattern that was seen
previously in CS 22892--052 and HD~115444 \cite{cow99}. This strongly
suggests a robust process for the production of these heavier neutron-capture
elements. Very recently the europium isotopic abundance fractions have 
been determined for \bd17, CS~22892--052 and HD~115444
\cite{sne02}. Those fractions 
are all in excellent agreement with each other and with their values in
the solar system. Thus, there appears to be 
a consistency  in the production
of the heavier neutron-capture  elements and, at least for europium, 
the isotopes.

We also note the trend of the lighter neutron capture elements in Figure 2.
Similarly to the case of CS~22892--052,  the elements with Z $=$ 40--50 
in \bd17 fall below  
the scaled solar system curve that fits the heavier 
neutron-capture elements. This might be explained by a single $r$-process site
with two regimes or sets of conditions, or
perhaps two different sites for
the lighter and heavier neutron-capture elements (see \cite{cow02}).

\begin{figure}[ht]
   \centerline{\epsfxsize=0.6\textwidth\epsffile{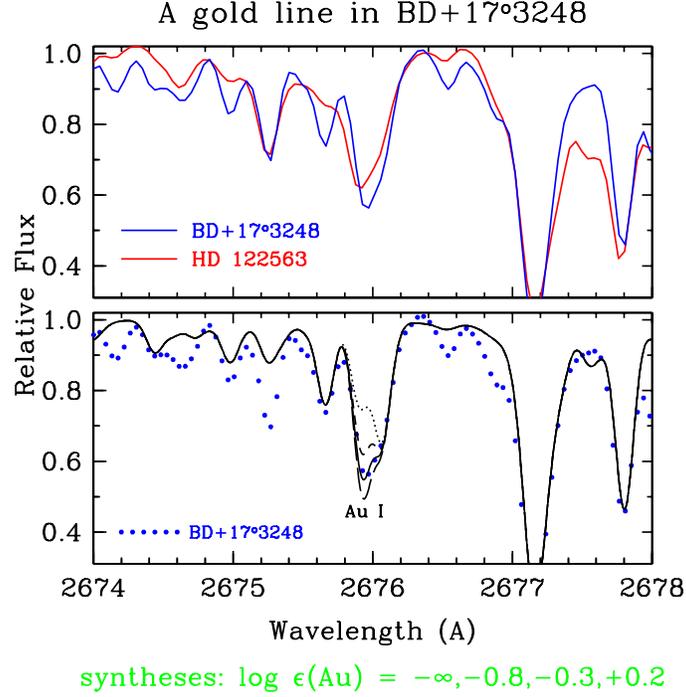}}
  \caption{
Observed HST-STIS and synthetic spectra in region surrounding the
Au I  2675.94~\AA\ line.
In the top panel, the observed \bd17\ spectrum is compared to that
of HD~122563.
In the bottom panel, the 
observed \bd17\ spectrum  is compared to four
synthetic spectra, shown in order of increasing abundance 
by dotted, short dashed, solid, and long dashed lines  
computed for log~$\epsilon$(Au)~= --$\infty$, --0.8, --0.3, +0.2.}
 \label{cowan.fig1}
\end{figure}

\begin{figure}[ht]
   \centerline{\epsfxsize=0.6\textwidth\epsffile{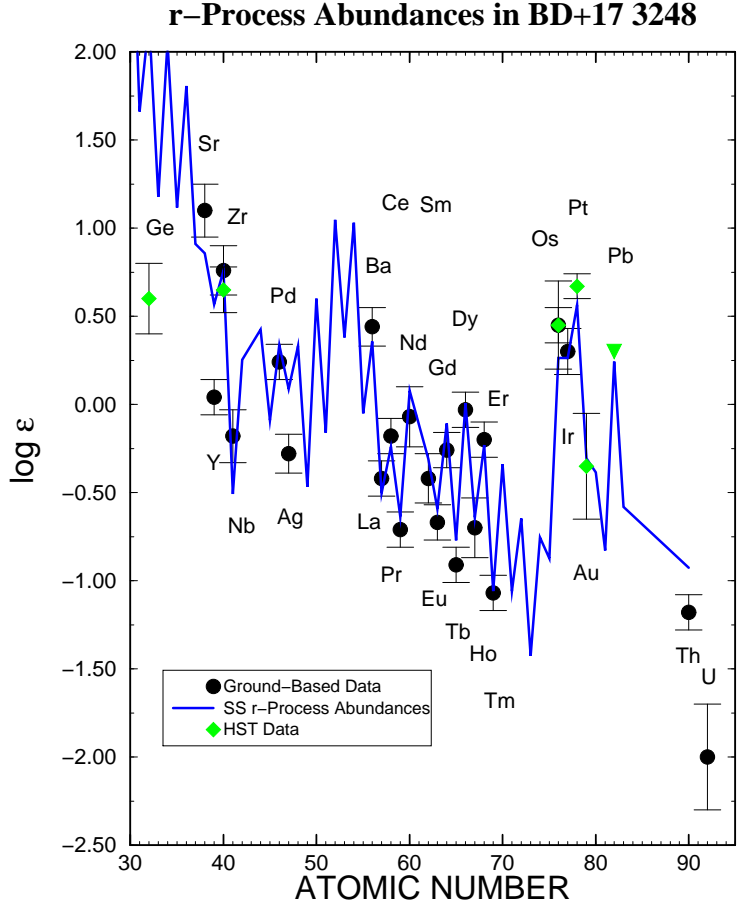}}
  \caption{
Neutron-capture element abundances in \bd17, obtained by ground-based
and HST observations, compared to a scaled solar system $r$-process
abundance curve.
The upper limit on the lead abundance is denoted by an inverted triangle.
Note also the thorium and uranium detections.}
  \label{cowan.fig2}
\end{figure}

\section{r-Process Chronometers}

The long-lived radioactive nuclei, such as $^{232}$Th and $^{238}$U, 
can be used as chronometers (or clocks) to determine the ages of stars.
Thus, the detection of thorium in a number of halo stars has led to 
stellar radioactive age estimates. To minimize observational 
errors the ratio of the 
radioactive element thorium (formed entirely in the $r$-process) is normally
compared to the stable ($r$-process) element europium. The detection of 
\third\ $r$-process peak elements such as Pt provides additional 
chronometric pairs, with the advantage that these elements are
closer in atomic mass (than Eu) to Th.  
Furthermore,
new stellar detections of U can  provide a second
chronometer
and help constrain age determinations.
Employing the newly detected Th, U and \third\ $r$-process peak
element abundances Cowan {\it et al.}\cite{cow02}  made chronometric 
age estimates for \bd17.
The average value of the various
chronometric pairs suggests an age of 13.8 $\pm$ 4 Gyr for this star.
This age estimate  is consistent, within error limits, with other chronometric
age determinations for  metal-poor Galactic halo stars.

The still relatively large uncertainties 
reflect  
the sensitivity of
the age estimates to  both the
observed and predicted (initial) abundance ratios.
Additional stellar observations will help to reduce observational
uncertainties. Even in cases where only upper limits on uranium 
may be  avaialable (perhaps the norm for most stars), 
the upper limits to the U/Th ratio can
already  provide lower limits on
the age estimates for the most metal-poor stars and hence constrain age
determinations for the Galaxy \cite{bur02}.
Additional theoretical studies, see   
\cite{bur02,sch02}, will also help to reduce the errors in the 
theoretical predictions for the initial abundances of the radioactive 
elements produced in the $r$-process.
Despite any  current uncertainties, 
this radioactive dating technique offers promise. It is
independent of chemical evolution or cosmological models. It offers 
an alternative  means of placing lower lmits on the ages of the Galaxy and
the Universe.

\section*{Acknowledgements}

We thank our colleagues and collabortors for many helpful discussions.
This research has been supported in part by STScI grants GO-8111 and GO-08342,
NSF grants
AST-9986974  (JJC),
AST-9987162  (CS), and by the ASCI/Alliances Center for Astrophysical 
Thermonuclear Flashes
under DOE contract B341495 (JWT).



\bbib
\bibitem{sne96} C. Sneden, {\it et al.}, ApJ {\bf 467} (1996) 819. 
\bibitem{sne00} C. Sneden, {\it et al.}, ApJ {\bf 533} (2000)  L139.
\bibitem{wes00} J. Westin, {\it et al.}, ApJ {\bf 530} (2000) 783.
\bibitem{cow02} J. Cowan, {\it et al.}, ApJ  (2002)  in press.
\bibitem{cay01} R. Cayrel, {\it et al.}, Nature {\bf 409} (2001) 691.
\bibitem{cow99} J. Cowan, {\it et al.}, ApJ {\bf 521} (1999) 194.
\bibitem{sne02} C. Sneden, {\it et al.}, ApJ (2002) in press.
\bibitem{bur02} S. Burles, {\it et al.}, ApJ (2002) in preparation.
\bibitem{sch02} H. Schatz, {\it et al.}, ApJ (2002) submitted.
\ebib
\end{document}